\documentclass{article}

\usepackage{arxiv}

\usepackage[utf8]{inputenc} 
\usepackage[T1]{fontenc}    
\usepackage{hyperref}       
\usepackage{url}            
\usepackage{booktabs}       
\usepackage{amsfonts}       
\usepackage{nicefrac}       
\usepackage{microtype}      
\usepackage{lipsum}
\usepackage{graphicx}
\usepackage{subfig}
\usepackage{mathrsfs}

\graphicspath{{images/}}

\title{Difficulty Scaling in Proof of Work for Decentralized Problem Solving}

\author{Pericles Philippopoulos \thanks{P. Philippopoulos (pericles.philippopoulos@protonmail.com) is co-founder of \={O}zeki Inc. a Blockchain Consultancy Firm in Montreal, Quebec, Canada}\and{Alessandro Ricottone \thanks{A. Ricottone is a Doctoral Student in Quantum Computing at McGill University, Montreal, Quebec, Canada}}\and Carlos G. Oliver \thanks{C. G. Oliver (cgoliver@protonmail.com, corresponding author) is co-founder of \={O}zeki Inc.}}

\begin{document}
\maketitle

\begin{abstract}
We propose DIPS ({\bf D}ifficulty-based {\bf I}ncentives for {\bf P}roblem {\bf S}olving), a simple modification of the Bitcoin proof-of-work algorithm that rewards blockchain miners for solving optimization problems of scientific interest. 
The result is a blockchain which redirects some of the computational resources invested in hash-based mining towards scientific computation, effectively reducing the amount of energy `wasted' on mining.
DIPS builds the solving incentive directly in the proof-of-work by providing a reduction in block hashing difficulty when optimization improvements are found.
A key advantage of this scheme is that decentralization is preserved and no additional protocol layers are required on top of the standard blockchain.
We study two incentivization schemes and provide simulation results showing that DIPS is able to reduce the amount of hash-power used in the network while generating solutions to optimization problems.
\end{abstract}


\section{Introduction}

Energy spent hashing in Proof of Work (PoW) is critical for guaranteeing the integrity of transactions on the ledger.
However, the information that results from the computations in itself is of low value.
More specifically, the outcome of mining a block is simply a proof that  there exists a nonce for which the hash of a given block is lower than some threshold.
Naturally, this knowledge is not applicable outside of the blockchain itself.
Mining protocols which secure the network and are themselves informative have therefore been an attractive goal since the early days of blockchain \cite{king2013primecoin,foldingcoin,ileri2016coinami}.
Meanwhile, crowd sourcing efforts have been successful in showing that solutions to difficult scientific problems can be discovered by a large community. ~\cite{kawrykow2012phylo,larson2009folding,anderson2002seti}
However, current crowd-sourcing models offer few incentives for participation which limits the size of the user-base.
On the other hand, cryptocurrency mining has been shown to offer a very strong incentive scheme and currently draws a much larger pool of contributors.
Including scientific computing in the mining protocol of a blockchain (more specifically a cryptocurrency) would therefore introduce and incentivize a larger community to solving scientific problems.

\subsection{Related Work}

We borrow the term `useful work'~\cite{ball2017proofs} to describe protocols which aim at incentivizing computations with real-world applications other than securing blockchain integrity, while acknowledging that this is not to say that standard PoW is not `useful'.
An obvious application of computational resources to `useful work' is finding solutions to problems of scientific interest such as NP-Complete problems or optimization tasks.

`Useful work' protocols can be grouped in two major types: {\it economy-based} and {\it mining-based}. 
The former has been proposed in works such as CureCoin~\cite{foldingcoin}, and Coinami ~\cite{ileri2016coinami} which use the interest in cryptocurrencies to reward users who complete certain tasks such as DNA alignments or protein folding directly with tokens.
However, for the most part the blockchain protocol remains untouched and PoW is still required in its entirety, typically with some additional centralization.
On the other hand, {\it mining-based} approaches attempt to replace PoW with alternative forms of work. 
Proposed methods can vary widely in this category.
Protocols such as PrimeCoin ~\cite{king2013primecoin} and Conquering Generals ~\cite{loe2018conquering} attempt to fully replace the classical PoW problem of block hashing to another NP-Complete problem.
These problems are picked such that they preserve the following properties:

\begin{enumerate}
    \item Solving is difficult (hash functions are non-invertible)
    \item Difficulty can be easily tuned (target value of hash digest determines difficulty)
    \item Fast solution verification (computing and comparing hashes is constant time)
    \item Easy to generate new problems (block data defines a problem)

\end{enumerate}

However, very few problems of scientific interest can satisfy all of these criteria.
The first protocol which supports arbitrary NP-Complete problems addresses these constraints by retaining a portion hash-based PoW ~\cite{oliver2017proposal}.
Miners optionally submit a block at a reduced {\it hashing} difficulty if the block includes a valid solution to a problem selected by the network.
If a solution to the problem is not found, the protocol converges to Bitcoin.
The difficulty of the problem is estimated by the frequency of blocks mined with solutions and difficulty is adjusted accordingly to maintain a desired block time.
Recently, ~\cite{chatterjee2019hybrid} proposed a special case of ~\cite{oliver2017proposal} which allows problem solvers to submit blocks without hashing (reduced difficulty of zero) if they provide a solution.
Attacks where malicious miners pre-solve many problems to win a large portion of consecutive blocks are prevented by enforcing that at least \%50 of blocks be mined classically (effective reduced difficulty half of classical difficulty).
Finally, ~\cite{amar2019incentive} rewards solutions to `useful' problems with votes in a hybrid Proof of Work/Proof of Stake model which, by definition, reduces the required amount of hashing.
However, this also comes with a complex additional layer of governance and potential centralization.

\subsection{Contribution}

Here we build on ~\cite{oliver2017proposal} to provide stronger problem-solving incentives while maintaining simplicity, full decentralization, and blockchain security.
We describe in  Section \ref{sec:protocol} a novel difficulty adjustment scheme which provides stronger problem solving incentives.
In Section \ref{sec:sim} we present simulations of the resulting network behaviour.
Finally, in Section \ref{sec:Bubka} we address potential attacks.

\section{Protocol}\label{sec:protocol}

Here we describe the DIPS protocol to incentivize the mining network to solve an agreed upon optimization problem through mining difficulty reductions.
For simplicity, we phrase an optimization problem $\mathscr{P}$ as a sequence of NP-Complete decision problems of the form `does there exist a solution with objective value greater than some fixed target?'.
As an example, $\mathscr{P}$ can be a specific graph for which we want to find the maximal clique.
Since verifying that a clique is maximal is NP-Hard, we let the network solve a series of decision problems by proposing larger and larger cliques.
Solutions to the decision problems can be checked in constant or linear time (is this a clique of size $k$?).

\subsection{Single update}\label{sec:v1}

In the first version ($v1$) of the protocol (see Ref.~\cite{oliver2017proposal}), miners are given the option of mining blocks `classically' (as in the Bitcoin protocol, Ref.~\cite{nakamoto2008bitcoin}) with a given difficulty $d_b$ or by including solutions to a given $\mathscr{P}$ in a block and mining that block with a reduced difficulty $d_r$.
Since there are two difficulties in $v1$, two conditions are required to be satisfied to update the difficulties. 
This situation is in contrast to the Bitcoin protocol, where a single difficulty is updated using a single condition: the average time required to mine each block $T$ is fixed (taken to be $ 10 \, ~\mathrm{minutes}$ in Bitcoin). 
The first condition used to update the difficulty in $v1$ is the same as in the Bitcoin protocol.
In addition to this condition, the difficulties are updated so that the average ratio between $d_r$ and $d_b$, $\eta$ is fixed,
\begin{equation}
    \left<\frac{d_r}{d_b}\right> = \eta,
\end{equation}
where $\left< Q \right>$ represents the long-time average of the quantity $Q$.
As is the case in Bitcoin, the difficulties are updated after $N_1$ blocks are mined ($N_1 = 2016$ in Bitcoin).
In $v_1$ both difficulties are updated independent of how many blocks are mined including a solution to $\mathscr{P}$.
Moreover, as is the case with Bitcoin, measures are taken to ensure that difficulty does not change too quickly.
In particular, the protocol enforces a maximal update factor for the difficulties so that 
\begin{equation}
     \frac{1}{x} \le \frac{d_i^{j+1}}{d_i^j} \le x,
\end{equation}
where $i \in \{b, r\}$, $d_i^j$ is the value for $d_i$ after $j N_1$ blocks and $x$ is the maximum factor by which the difficulties can be updated. 
After $j N_1$ blocks, if the difficulty $d_i^{j+1}$ is calculated so that $x < d_i^{j+1}/d_i^j$ ($ d_i^{j+1}/d_i^j < 1/x$) then we take $d_i^{j+1} = x d_i^{j} (d_i^{j+1} = d_i^{j} / x )$ instead.

Although this version of the protocol incentivizes solving NP-complete problems by providing a reduced mining difficulty (assuming $\eta < 1 $), $d_r$ increases with $d_b$ since the average value of $d_r/d_b$ is fixed. 
Therefore, even in the case where no solutions are submitted, $d_r$ can increase.
This means that even as $\mathscr{P}$ becomes more difficult to solve (which is expected to happen over time), $d_r$ can continue to increase limiting the incentive provided by the network to solve problems and therefore the number of solutions that the network will find.
Consequently, the amount of resources redirected to solving useful problems will also be limited. 
    
\subsection{Independent updates}\label{sec:v2}
We propose a second version of the protocol ($v2$) where miners are again free to submit `classical' blocks with difficulty $d_b$ or blocks containing a solution to an NP-complete problem, with difficulty $d_r$.
However, in $v2$ the difficulties $d_b$ and $d_r$ are updated independently. 
That is, after $N_2^b$ classical blocks have been mined, $d_b$ is updated so as to keep the average time spent by the network to mine a classical block fixed (to a predetermined value $t_2^b$). 
Similarly, after $N_2^r$ blocks have been mined containing a solution, $d_r$ is updated so as to keep the average time spent by the network to mine a block with a solution be fixed (to a predetermined value $t_2^r$).
Since it becomes increasingly difficult to find solutions to NP-complete problems, eventually the network might not be able to mine new blocks with the current difficulty $d_r$. 
Therefore in the $v2$ protocol if $N_2^b$ consecutive classical blocks are mined, $d_r$ is decreased by the maximum factor, $x$ ($d_r \rightarrow d_r/x$).
In this way, even if the problem becomes increasingly difficult to solve, miners that attempt to solve the problem are incentivized with a proportionally decreased $d_r$. 
Miners are naturally discouraged from holding on to their solution until the difficulty is lowered by a large factor since this comes at the risk of other miners finding and publishing a better solution first.

\begin{figure}
\centering
\includegraphics[width=.8\textwidth]{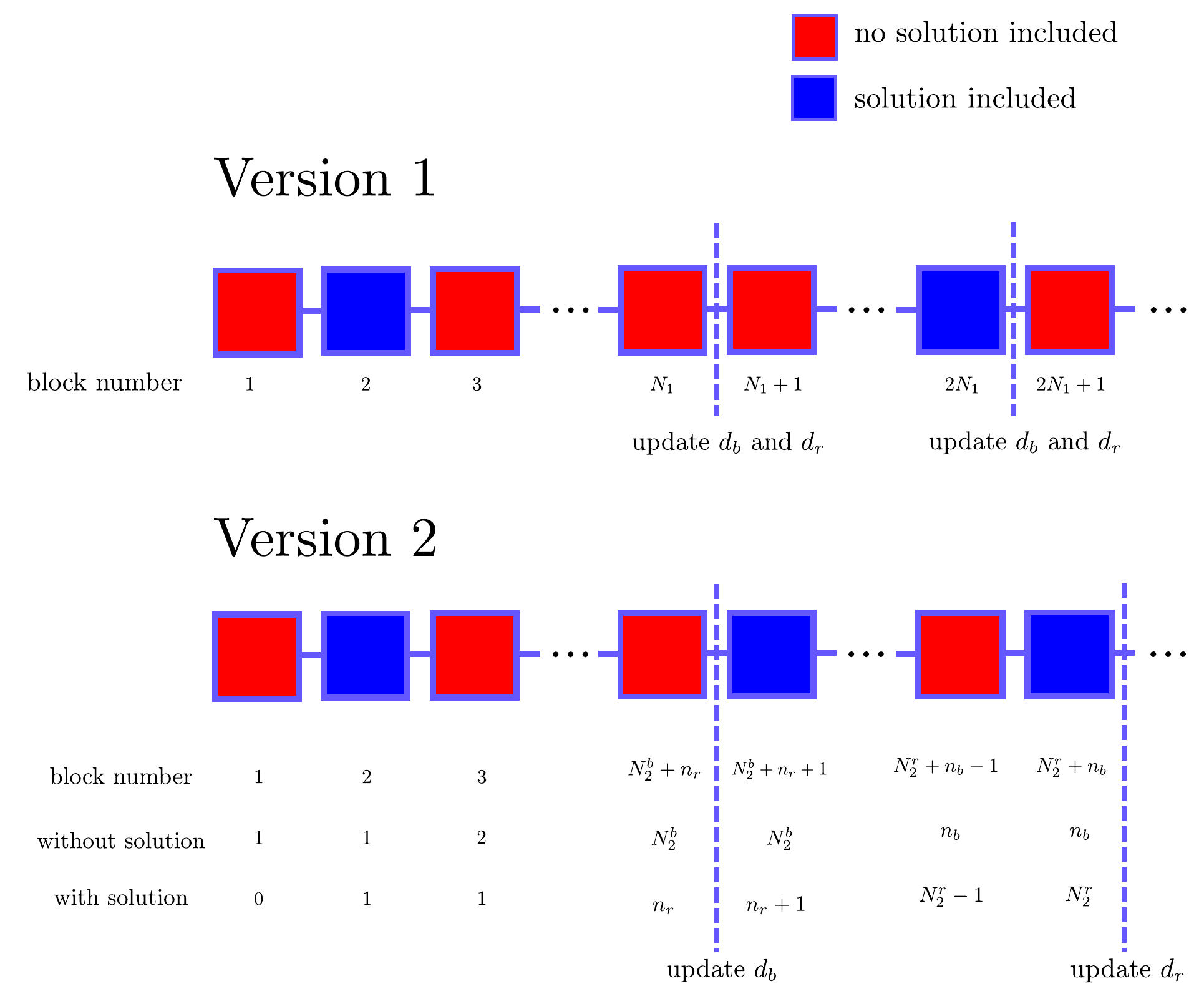}
\caption{Schematic depicting the difference between a blockchain with proof-of-work protocol $v1$ and $v2$.
The red blocks are standard blocks that do not contain a solution to a problem (classical blocks), while the blue blocks contain a solution that results in a score that is better than the current best score of $\mathscr{P}$ (solution blocks). 
In $v1$ both $d_b$ and $d_r$ are updated every $N_1$ blocks. 
In $v_2$, $d_b$ is updated every $N_2^b$ blocks that do not contain a solution to a problem and $d_r$ is updated every $N_2^r$ blocks that do contain a solution. 
$n_r$ and $n_b$ are arbitrary numbers. }\label{fig:versions}
\end{figure}

\subsection{Problem Submission}\label{sec:newprob}
In the type of modified proof-of-work protocol described in the above subsections, some of the hashing power reserved in traditional blockchains (such as Bitcoin) to ensure the security of the network is diverted to solving an NP-complete problem, $\mathscr{P}$. 
The difficulty in finding better and better solutions provides the network with security and replaces the repeated hashing of proposed blocks. 
Because the best score of $\mathscr{P}$ will eventually saturate (either when the network no longer wishes to find a better score or when the optimal score has been found) this modified proof-of-work system will revert back to the traditional proof-of-work protocol.
Therefore, to create a blockchain where NP-complete problems keep being solved, it might be interesting to consider a system where new problems can be submitted to the network. 

The inclusion of new problems in the network can be done in a variety of ways.
We leave determining a specific implementation for a future work and discuss some possible implementations here instead. 
One possible implementation could have a committee of users or a group of special nodes chosen (through an election, or otherwise) to submit new problems once the best score of the current problem has been saturated. 
Alternatively, but in the same spirit, individuals (participating in the network, or not) can propose new problems off-chain and the community can vote on which new problem to replace the current problem with.
Once the new problem has been chosen, the blockchain can be hard forked to ignore solutions to the old (saturated problem) and accept solutions to the new agreed-upon problem. 

A different idea for submitting problems would allow any node to submit a problem as long as they stake some number of tokens (native to the blockchain) on the problem being submitted. 
This stake is `burned' by the network.
Otherwise, the party submitting the problem could gain free tokens by submitting an easy problem they already have solutions to and subsequently mining a block at the reduced difficulty and gaining the mining reward.
This type of implementation can be dangerous because if the number of tokens the network requires submitters to stake is too low, new solved problems will constantly be submitted so that the submitters can mine new blocks with a reduced difficulty for free.
Moreover, submitters can hoard multiple solutions to the same problem that can be used to gain a difficulty reduction in mining for many consecutive blocks (see section \ref{sec:Bubka} for more details on this type of attack).
In contrast, if the stake required is too high, many users will be discouraged to submit interesting problems. 
    
\section{Simulation Results}\label{sec:sim}

Here we perform simple experiments to visualize the impact of protocol and parameter choice on the behaviour of the network.
We implement a simulated blockchain with 10 miners performing classical mining and 10 miners problem solving.
As a sample problem, we let the network solve the well-known maximum clique problem on randomly generated graphs ~\cite{tomita2006worst}.
At the start of the chain, we generate a random graph and miners apply the Bronn-Kerbosch ~\cite{bron1973algorithm} algorithm which enumerates cliques until a large enough solution is found.
The current largest clique size is stored in each block.
Since mining hardware is typically used only for hashing, we assume miners are concurrently mining at $d_b$ while attempting to solve the problem on separate hardware.
If a miner finds a clique which beats the current best, he includes the solution in his next block and begins to mine at a reduced difficulty $d_r$.
If he fails to win the next block, the miner keeps his current best solution to try again in the next block.

\begin{figure}
    \includegraphics[width=\columnwidth]{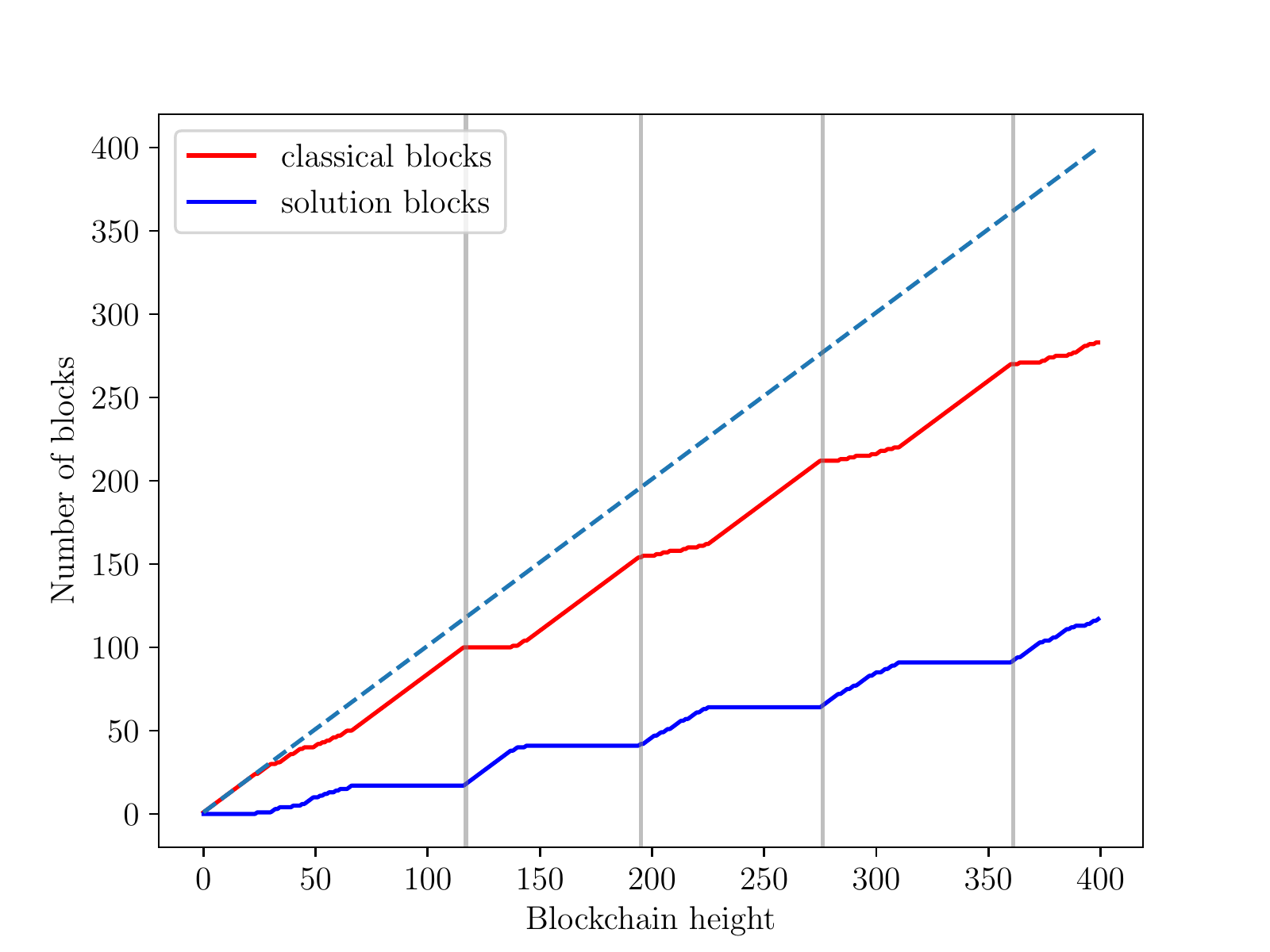}
    \caption{Number of classical (red) and solution blocks (blue) mined as a function of the blockchain height using the simulations described in section~\ref{sec:sim}.
    The dashed line represents the number of classical blocks mined on a classical blockchain (such as Bitcoin) as a function of the blockchain height. 
    We have also replaced the problem in the network by a new one once the best score saturates. 
    The introduction of a new problem is represented by the gray vertical lines. We use protocol $v2$ with $t_2^b=t_2^r=0.1s$. The update frequency for solution blocks is set to $N_2^b=10$ and $N_2^r = 5$ since at early stages of the problem, solutions are found quickly. 
    }
    \label{fig:bcheight}
\end{figure}

In Figure~\ref{fig:bcheight} we plot the number of classical and solution blocks as a function of the blockchain height. 
In a standard proof-of-work blockchain (such as Bitcoin), there is only one type of block - a classical block. 
The dashed line in Figure~\ref{fig:bcheight} represents how the number of classical blocks grows with blockchain height in a standard blockchain (they grow together since they represent the same quantity). 
In contrast, for the $v2$ proof-of-work protocol, some of the hashing power is diverted to solving NP-Complete problems. 
Therefore there are fewer fully classical blocks mined at a given blockchain height and therefore, less energy invested by the network to hash blocks.
The energy saved is used to solve NP-Complete problems as indicated by the blue curve, which represents blocks being mined by the network that contain solutions to NP-Complete problems - solution blocks. 

A feature of optimization problems is that eventually the optimal solution to the problem will be found. 
At this point (or before) the number of blocks with solutions will saturate because no new solutions can be found. 
The system reverts to a standard proof-of-work blockchain. 
One way to prevent the system from becoming a standard blockchain is to introduce new problems when the solution saturates (see section~\ref{sec:newprob} for a discussion on possible ways to include new problems).
We have simulated the inclusion of a new problem (randomly generated graph) when the best score of the previous problem saturates (gray vertical lines in Fig.~\ref{fig:bcheight}).
In this way, instead of reaching a point where no more energy is diverted to solving NP-Complete problems, as the blockchain grows, more energy is diverted to solving these types of problems. 
This energy is loosely represented by the difference in the dashed line and the red curve in Fig.~\ref{fig:bcheight}, or equivalently, the blue curve. 

\begin{figure}
  \centering
  \subfloat[$v1$]{\label{fig:subfig1}\includegraphics[clip=true, width=0.44\textwidth]{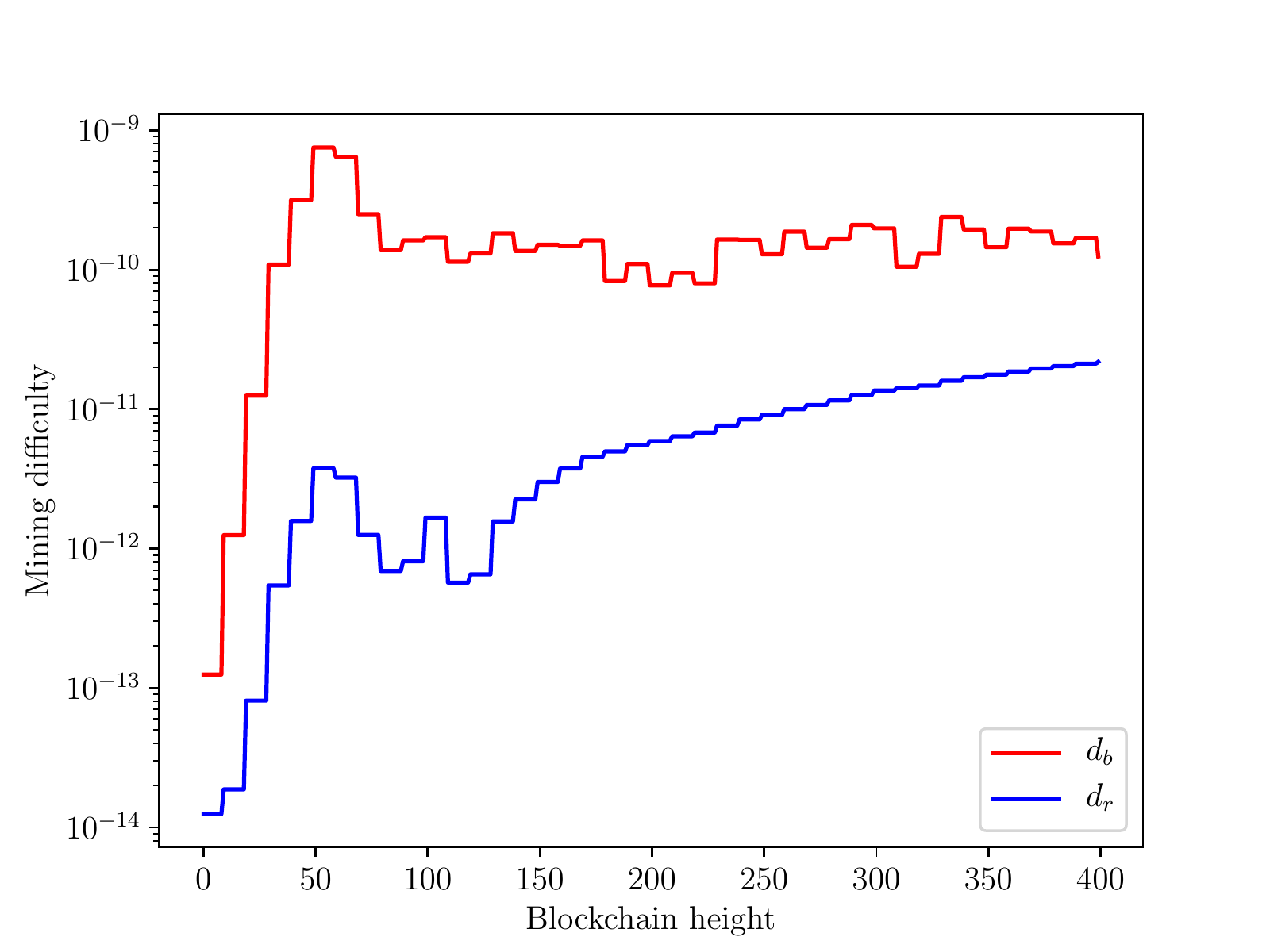}} \quad
  \subfloat[$v2$]{\label{fig:subfig2}\includegraphics[clip=true,  width=0.44\textwidth]{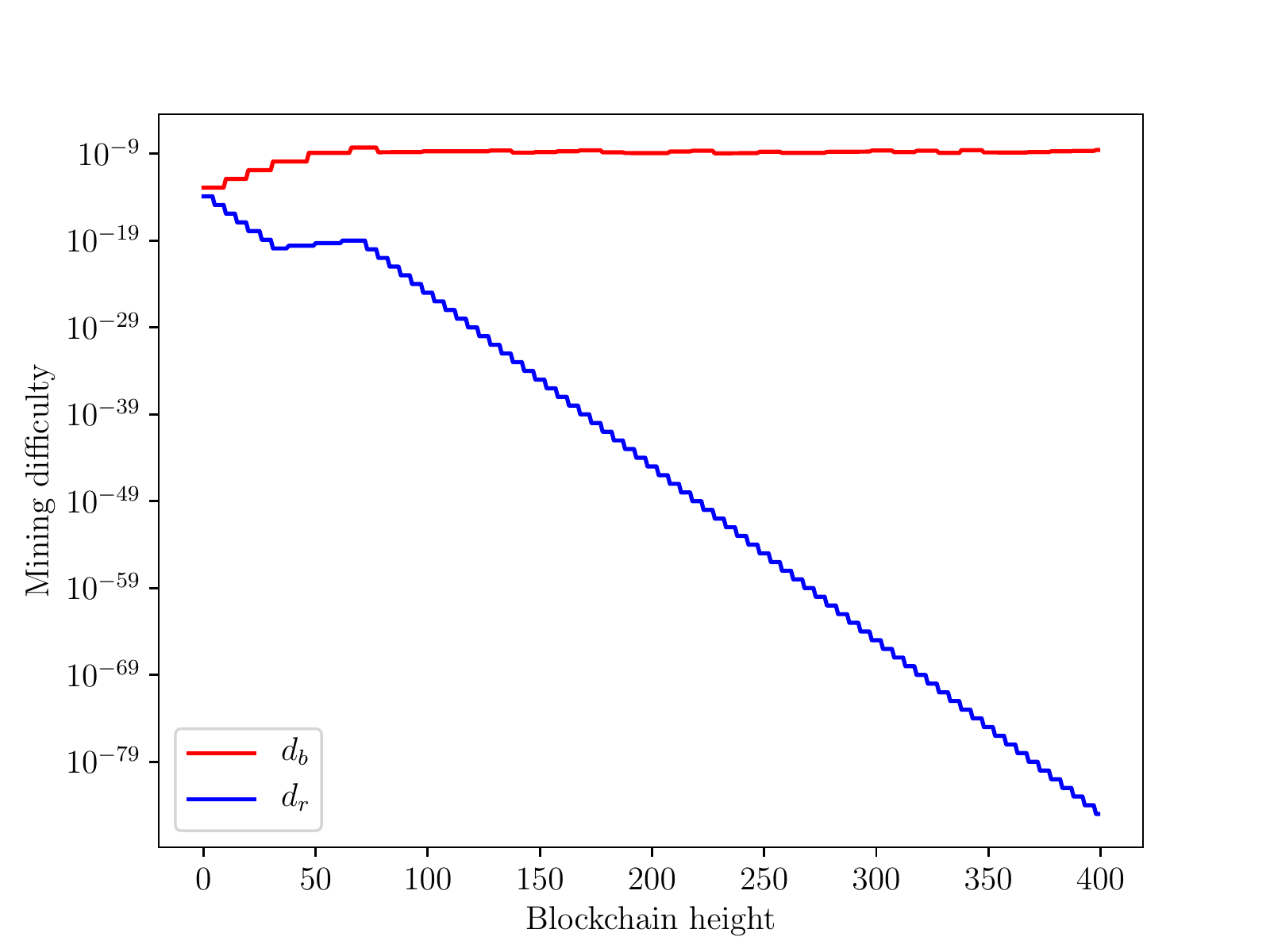}}

  \caption{Difficulties $d_r$ (blue) and $d_b$ (red) as a function of block height for a typical run of the simulation described in section~\ref{sec:sim}. For $v1$ we set $\eta = 200^{-1}$, and remaining parameters are kept from Fig. \ref{fig:bcheight}}
  \label{fig:diffs}

\end{figure}

\begin{figure}
    \includegraphics[width=\columnwidth]{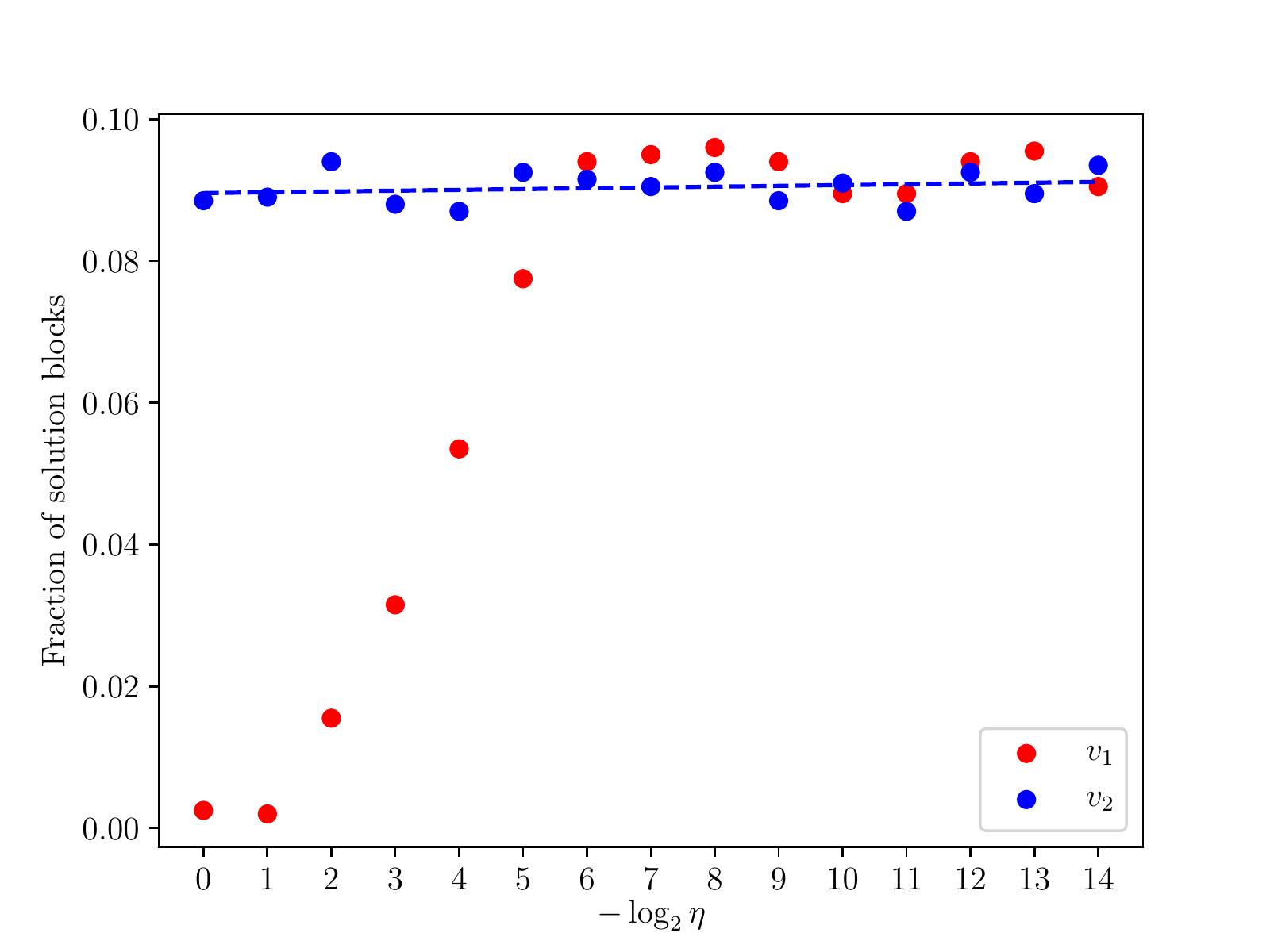}
    \caption{Fraction of blocks mined with a solution after 200 total blocks for the proof-of-work protocol $v1$ (red) and $v2$ (blue) as a function of $\eta$.
    In $v1$, $\eta$ is enforced by the network, while in $v2$ it is the initial ratio $d_r/d_b$.
    The fraction of blocks with solutions does not depend on the value of $\eta$ for $v2$. 
    The blue dashed line is the average value of the fraction of blocks with solutions over values of $\eta$.
    For $v1$ the fraction of solution blocks converges to the $v2$ value (dashed blue line) in the limit where $\eta \rightarrow 0$. Each point represents an average of 10 independent blockchain instances of height 200. Parameters are chosen as in Fig. ~\ref{fig:versions}}
     \label{fig:fraction}
\end{figure}

We have also studied how this energy is affected by the parameter $\eta$.  
For the $v1$ protocol, $\eta$ is a parameter that is enforced by the network as the difficulties get updated (see section~\ref{sec:v1}).
In the case of the $v2$ protocol, $\eta$ indicates the initial value of $d_r/d_b$.
As expected, since $\eta$ is not enforced by the network, the fraction of blocks that are solution blocks is independent of $\eta$ for $v2$ (see Fig.~\ref{fig:fraction}).
In contrast, in $v1$, $\eta$ determines how much incentive the community is given to solve the blockchain's optimization problem.
Therefore as $\eta$ decreases (as you move to the right along the horizontal axis of Fig.~\ref{fig:fraction}) we expect the fraction of solution blocks to increase as is shown in Fig.~\ref{fig:fraction}.
Furthermore, because the network enforces that the ratio $d_r/d_b$ (long-time) averages to $\eta$, $d_b$ and $d_r$ vary together (see Fig.~\ref{fig:subfig1}).

After a certain time, as the problem becomes increasingly more difficult to solve, $d_r$ will not decrease (and might even increase if the hashrate of the network increases) if the ratio $d_r/d_b$ has reached its equilibrium value $\eta$.
Therefore in $v1$ a situation is eventually reached where the problem difficulty increases and $d_r$ remains fixed i.e. as the problem becomes more difficult, the hashing difficulty remains constant, increasing the total difficulty of mining a solution block.
This type of situation does not encourage nodes to solve problems. 
Alternatively, in $v2$ if no solutions are found, $d_r$ decreases (see section~\ref{sec:v2} and Fig.~\ref{fig:subfig2}).
Therefore, as the problem becomes more difficult and new solutions become harder to find, the incentive for finding new solutions increases.
These aspects of the protocol are reflected in Fig.~\ref{fig:fraction}.
We find that for $v2$ the best score always saturates, while for $v1$ the saturation occurs only if $\eta$ is small enough. 
As $\eta$ decreases, the fraction of blocks that are solution blocks in $v1$ tends to the saturated value (average $v2$ value, given by the dashed blue line).

\section{Potential Attacks}\label{sec:Bubka}
 
 Besides the standard attacks possible with traditional proof-of-work blockchains (e.g. 51\% attacks performed by controlling a majority of the network's mining power), the DIPS proof of work introduces additional attack avenues. 
We name the main attack, to our knowledge, for this version of proof of work the {\it Bubka attack} after Ukrainian pole vaulter Sergey Bubka who obtained many consecutive world records by incrementally improving his score instead of posting one world record by a large margin.
 This attack strategy involves finding multiple successive solutions to  $\mathscr{P}$ and using these to get an advantage in hashing for multiple blocks in a row. 
  Worse still, one could copy the solutions to problems from solution blocks and use them to fork the chain at a different block height.
  This strategy would allow the attacker to double spend by forking the network at some past block and creating the longest chain with less than 51\% of the network's hashing power. 
  Estimating the exact fraction of hashing power required for a double-spend attack in this case is not straightforward and depends on, among other factors, the ratio $d_r/d_b$ at the time of the attack, the rate at which the network finds new solutions to the problem, the hash rate of the network, and the values of the parameters $N_2^r$ and $N_2^b$.
  We leave the exploration of this question for future work.
  
 Other attacks such as well-connected miners claiming other miners solutions as their own are addressed in ~\cite{amar2019incentive} with protocols that are fully compatible with DIPS.

 There are several options for addressing this threat on the social layer of the network.
 Unlike in other proposed methods, mining a solution to a block still requires a hashing step, the network could therefore choose a large enough update frequency (small enough value of $N_2^r$) to rapidly increase mining difficulty if solutions are posted in rapid succession.
 Additionally, users could follow higher transaction `confirmation times' or not allow long consecutive chains of blocks with solutions to the problem.

\section{Conclusion}
In this article we have compared two modified proof-of-work protocols that divert energy from mining by hashing blocks to solving NP-Complete problems (useful work). 
We have studied, through simulations, the dependence on the total diverted energy on different parameters.

Although this article provides a protocol that can convert the energy used by miners to a form of useful work (solving NP-Complete problems), creating a fully functioning and useful blockchain using any one of the two protocols ($v1$ or $v2$) would require the understanding of other aspects of system.
For example, a concrete protocol to add new problems to the network must be developed.
Moreover, a standard way of storing and solving submitted problems must be established.
Other works have made some effort to solve these problems\cite{amar2019incentive,chatterjee2019hybrid}, however more work is needed to adapt these solutions for use in the proof-of-work protocols $v1$ and $v2$. 
We believe protocols such as DIPS are an important step towards combining the potential of crowd-sourcing initiatives (such as Folding@Home ~\cite{larson2009folding} and Phylo ~\cite{kawrykow2012phylo} which have  resulted in new solutions for important problems) with the strong incentive structures native to blockchains.


\section*{Author Contributions}

P.P and C.G.O conceived the research, prepared the manuscript and performed the simulations. A.R. helped conceive the simulation protocol.

\clearpage
\bibliographystyle{unsrt}
\bibliography{references}

\end{document}